# Nanosecond compressive fluorescence lifetime microscopy imaging via the RATS method with a direct reconstruction of lifetime maps

**JIŘÍ JUNEK**[1,2,*] **AND KAREL ŽÍDEK**[1]

[1]*Regional Center for Special Optics and Optoelectronic Systems TOPTEC, Institute of Plasma Physics of the Czech Academy of Sciences, Za Slovankou 1782/3, 182 00 Prague 8, Czech Republic*
[2]*Technical University in Liberec, Faculty of Mechatronics, Informatics and Interdisciplinary Studies, Studentská 1402/2, 461 17 Liberec, Czech Republic*
*\*junekj@ipp.cas.cz*

**Abstract:** The RAndom Temporal Signals (RATS) method has proven to be a useful and versatile method for measuring photoluminescence (PL) dynamics and fluorescence lifetime imaging (FLIM). Here, we present two fundamental development steps in the method. First, we demonstrate that by using random digital laser modulation in RATS, it is possible to implement the measurement of PL dynamics with temporal resolution in units of nanoseconds. Secondly, we propose an alternative approach to evaluating FLIM measurements based on a single-pixel camera experiment. In contrast to the standard evaluation, which requires a lengthy iterative reconstruction of PL maps for each timepoint, here we use a limited set of predetermined PL lifetimes and calculate the amplitude maps corresponding to each lifetime. The alternative approach significantly saves post-processing time and, in addition, in a system with noise present, it shows better stability in terms of the accuracy of the FLIM spectrogram. Besides simulations that confirmed the functionality of the extension, we implemented the new advancements into a microscope optical setup for mapping PL dynamics on the micrometer scale. The presented principles were also verified experimentally by mapping a LuAG:Ce crystal surface.

## 1. Introduction

Fluorescence lifetime imaging (FLIM) is an important part of spectrometry dealing with photoluminescence (PL) dynamics. It is used mainly in biology [1,2], chemical physics [3,4], as well as in materials engineering [5-8].

There are several methods commonly used for FLIM. The best-known are time-correlated single-photon counting (TCSPC) [9], streak camera [10], gated photon counting [11], and the analog time-domain or frequency-domain technique [12,13]. However, each method has its principle limitations and, as a result, its preferred field of application. In other words, while a given time-resolved method is perfect for a certain use, it may be completely unsuitable for another [14].

Although FLIM has been developed for decades, new concepts which further improve the technique have been reported recently. The main goal is to obtain measurements and evaluations in real time, mainly to investigate biological and medical samples [15,16]. However, real-time FLIM acquisition depends on using pulsed lasers, digitizers, GPUs or 2D arrays of single photon avalanche diodes, which considerably increase the cost of such setups. This holds also for the new FLIM alternatives, such as phasor spectral FLIM (Phasor S-FLIM) [17]. Another competing concept, the analog mean delay method (ADM) [18], can be used for real-time vivo measurements. Nevertheless, ADM cannot be used to determine the actual shape of PL decay.

Therefore, FLIM research has also developed in the direction of using compressed sensing (CS), where the most typical application is the single-pixel camera (SPC) experiment, which makes the FLIM optical setup more affordable [19]. SPC experiments are very useful in terms of reducing the cost of the experimental setup, but they are typically not suited for real-time in-vivo imaging—both due to the sequential signal acquisition and consequent data retrieval. However, new post-processing approaches for SPC in FLIM are still being developed, e.g., the use of deep learning, which significantly shortens reconstruction time [20]. On the other hand, the neural network needs to be properly trained in order to obtain the required performance over the necessary time range of the investigated lifetimes.

Most of the improvements in CS are applied to standard time-resolved methods and thus still carry their principle limitations. E.g., TCSPC achieves excellent results for nanosecond or sub-nanosecond decays. However, for the measurement of decays on the order of hundreds of nanoseconds (field of material engineering), the method requires a measurement time of hours.

Therefore, it is still essential to continue developing new PL dynamics analysis methods. An example is the RAndom Temporal Signals (RATS) method, which is one of the novel approaches to FLIM. The method is based on the excitation of the measured sample via a randomly fluctuating intensity in time, which makes it possible to fully retrieve the PL decay from a single measured dataset. The RATS and 2D-RATS methods have proven to be valuable, straightforward, and low-cost alternatives to the commonly used FLIM approaches, although their previous implementation had limited their use in real-life experiments.

The initial proof-of-principle measurements of PL decay via RATS were based on a simple generator of a random excitation signal, where we focused a beam on a rotary diffuser and cropped the generated field of speckles with a suitable iris aperture [21]. Such a generator allows for the generation of an analog random excitation in the microsecond and sub-microsecond region, which is well suited for long-lived luminophores or nanoporous silicon [3,22]. However, reaching faster timescales is not realistic in this configuration.

At the same time, employing a diffuser limited the overall efficiency of the excitation laser because only a fraction of the diffused light was used. This fact was even more pronounced in our proof-of-principle implementation of the RATS method in FLIM (2D-RATS), which demonstrated that it is possible to map PL dynamics via an optical setup based on two diffusers [23].

Finally, the initially used post-processing consists of the reconstruction of PL maps at each delay after excitation and subsequent fitting to obtain a lifetime in a given image pixel. This procedure is highly computationally costly.

This article provides an overview of a new implementation of the 2D-RATS, i.e., the FLIM experiment, which overcomes all the above-stated shortcomings. This was enabled by three fundamental modifications in the RATS experiment. In particular, we introduced a new digital mode of random intensity fluctuation by using direct laser modulation. Fast laser modulation enabled us to reach the temporal resolution in the PL decay measurement of 6 ns. Secondly, using a digital micromirror device (DMD) to invoke random spatial masks, we gained incomparably higher efficiency reaching values a hundred times higher compared to the proof-of-principle experiments [23].

Finally, we propose a novel approach to PL data analysis where we first identify the significant PL lifetime components. An amplitude map of each component is subsequently computationally extracted. Therefore, a typical analysis—a bi-exponential fit of 35×35 map with a compression ratio of 0.4 and 100 timepoints of PL decay—leads to post-processing times about 10-times shorter compared to the approach used in our previous work [23].

Overall, the presented implementation shifts the abilities of the 2D-RATS method to a different level, which we demonstrate in both the synthetic and experimental data. The ability to retrieve PL dynamics on the scale of tens of nanoseconds in combination with the ability to cover long-lived PL decay make the method very useful in the field of material engineering [7,8]. At the same time, the current implementation remains a low-cost solution compared to standard FLIM setups.

## 2. Overview of the RATS method

### 2.1 0D-RATS

As mentioned in the introduction, the principle of the method is to excite the measured sample with a random excitation signal $I_{EXC}$. The generated photoluminescent $I_{PL}$ signal, which is a convolution of the excitation signal and the PL decay of the measured sample $I_D$, also has a random character as a result of the random excitation:

$$I_{PL} = I_{EXC} * I_D. \tag{1}$$

The PL decay $I_D$ curve can be calculated by using deconvolution. In the presented results, the Tikhonov regularization was applied to avoid ill-conditioned problems [24]:

$$I_D = Re\left\{\mathbb{F}^{-1}\left[\frac{\mathbb{F}(I_{PL})\,\mathbb{F}^*(I_{EXC})}{\mathbb{F}(I_{EXC})\mathbb{F}^*(I_{EXC}) + \varepsilon\overline{\mathbb{F}(I_{EXC})\mathbb{F}^*(I_{EXC})}}\right]\right\}. \tag{2}$$

Moreover, the Tikhonov regularization parameter $\varepsilon$ can also be used to suppress the noise effect of the system [25]. More details about a single-point measurement approach (0D-RATS) can also be found in our previous works [21, 23].

### 2.2 2D-RATS

In the case of 2D measurements, a straightforward option is to use the 0D-RATS method in raster mode, which can be very effective, for example, using a mirror with a Micro Electro Mechanical System (MEMS) [26]. Nevertheless, the use of compressive sampling in the configuration of the single-pixel camera ensures the acquisition of the desired dataset by a lower number of measurements compared to the raster mode, where the number of measurements corresponds to the number of pixels of an image [27].

The principle of the so-called 2D-RATS measurement consists in illuminating the scene with a set of $M$ random masks of $N$ pixels, where the ratio $M/N$ indicates the compression ratio $k$ [23]. The individual masks have a random character in space, and their intensity varies in time according to $I_{EXC}$—see Fig. 1(A). The randomness of the mask ensures that a different $n$ illuminated segment of the sample is excited in each measurement. Therefore, Eq. (1) can be rewritten to the sum of the PL from each segment:

$$I_{PL} = \sum_{i=1}^{n} I_{PL}(i) = I_{EXC} * \sum_{i=1}^{n} I_D(i). \tag{3}$$

As a result, we always attain a unique $I_{PL}$ for each of the $M$ masks. By using the $I_{PL}$ dataset, it is possible to reconstruct the individual $I_{DA}$ decay curves corresponding to the given mask:

$$I_{DA} = Re\left\{\mathbb{F}^{-1}\left[\frac{\mathbb{F}\left(\sum_{i=1}^{n} I_{PL(i)}\right)\mathbb{F}^*(I_{Exc})}{\mathbb{F}(I_{Exc})\mathbb{F}^*(I_{Exc}) + \varepsilon\overline{\mathbb{F}(I_{Exc})\mathbb{F}^*(I_{Exc})}}\right]\right\}. \tag{4}$$

Thus, each random mask is represented by one $I_{DA}$.

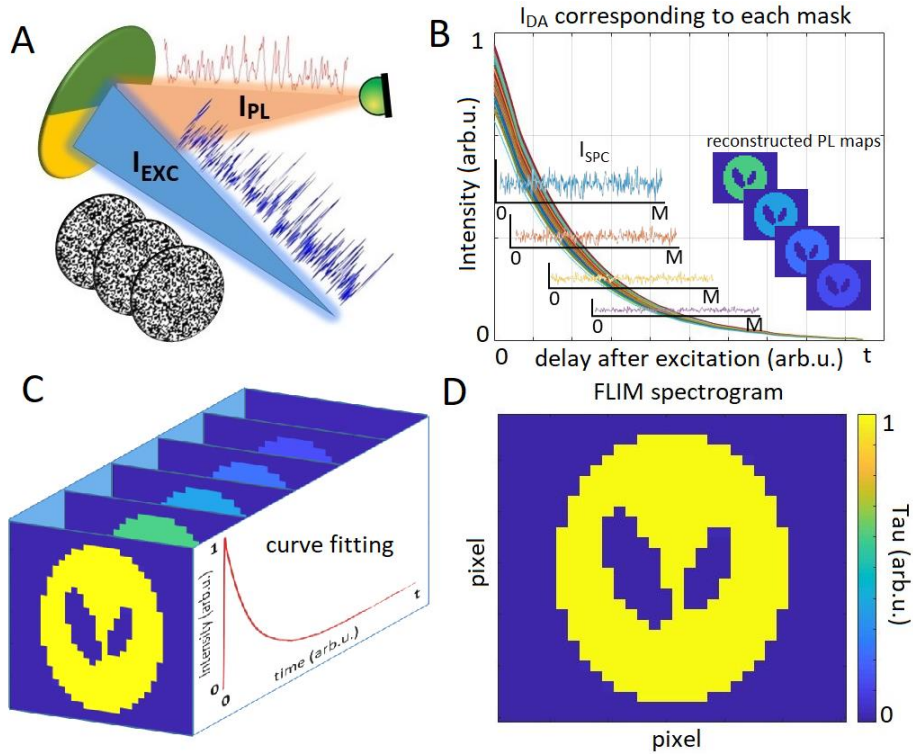

Fig. 1. (A) Scheme of principles of the 2D-RATS method. The sample is illuminated with masks following a temporal fluctuation of excitation signal $I_{EXC}$, which generate the corresponding PL signal $I_{PL}$. (B) Set of calculated $I_{DA}$ (solid lines in the main graph) corresponding to the set of masks can be recalculated into the $I_{SPC}$ signals (inset graphs for several delays $t$). By using Eq. (5) we attain the corresponding reconstructions of the PL maps. (C) Scheme of a 3D datacube of reconstructed PL maps and indicated fitting. (D) Example of a final FLIM spectrogram as a map of PL lifetimes.

The measured set of $I_{DA}$ curves then needs to be converted into the temporal slides of PL decay. To do this, we need to determine the PL intensity $I_{SPC}$ at a selected delay for all masks—see Fig. 1(B). The $I_{SPC}$ dataset is simply extracted from the $I_{DA}$ curves for the studied delay after excitation $t$. By using the knowledge of the random masks and the $I_{SPC}$, the PL map for each delay after excitation $m(x,y,t)$ can be determined using standard compressed sensing algorithms employed for the single-pixel camera experiment. We aim at solving an undetermined system using Eq. (5), where the set of vectorized masks is stated as $B$ and $TV$ is stated for total variation and the PL map for a given delay after excitation is declared just as $m(x', t)$ because it is vectorized:

$$min\left\{\|Bm(x',t) - I_{SPC}\|_2^2 + TV(m(x',t))\right\}. \quad (5)$$

The presented approach uses a well-established image retrieval procedure from the single-pixel camera experiment. A significant drawback is that this approach requires reconstructing PL maps for all delays after excitation and creating a 3D datacube to obtain complete information—see Fig. 1(C). This implies carrying out typically tens of iterative solutions of Eq. (5), which is computationally very costly. Finally, to attain the FLIM spectrogram (Fig. 1(D)), it is necessary to fit the third dimension (temporal dependence) of the 3D datacube in each pixel with a suitable function—typically a multi-exponential fit—and determine the lifetime $\tau$. For better understanding, refer to Fig. 1, which summarizes the whole procedure.

Compared to standard FLIM approaches, such as TCSPC or gated photon counting, 2D-RATS provides very robust PL retrieval, where the only parameter to be kept in mind is to ensure is that the setup has a sufficiently low width of the impulse response function (IRF) compared to the measured lifetimes, i.e., a sufficiently fast random signal. In contrast, the suitability of the setting parameters in TCSPC must be checked with respect to the repetition rate of excitation pulses or PL intensity. In the case of using gated photon counting, the user should choose the appropriate number of gates, length of gate, and other parameters.

### 3. Digital RATS modulation with direct PL lifetime map reconstruction

*3.1 Digital random signal generation*

The previously published proof-of-principle RATS experiment was based on generating a random excitation pattern by using a rotary diffuser. This leads to a rapidly fluctuating signal, which we denote as a random analog signal, as the light intensity continuously randomly varies in time. While this approach is useful owing to its simplicity and negligible cost, it limits the use of the RATS setup with respect to the reachable temporal resolution.

Therefore, we implemented an advanced mode of random signal generation with higher efficiency by using a modulated laser. One of the lasers that makes this possible is the Cobolt S06-01 (405 nm), which can be modulated digitally up to 150 MHz in a random fashion. The result is a rectangular signal with a randomly distributed duty cycle, which we denote as a random digital modulation.

We compare the digital and analog generation of a random signal in 0D-RATS on simulated data, while the experimental demonstration can be found in Section 4. An example of a simulated digital signal can be seen in Fig. 2 (upper left panel), where the character of the signal is more apparent in a zoomed part of the signal from 0 to 0.5 ms (see inset).

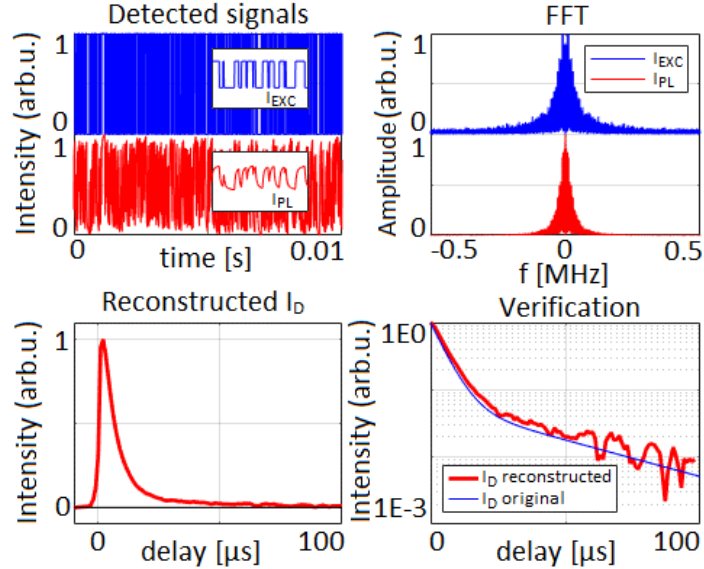

Fig. 2. Simulated measurement of bi-exponential $I_D$ with parameters $A_1 = 1$, $A_2 = 15$, $\tau_1 = 40$ μs, and $\tau_2 = 5$ μs using digital random signal with $IRF_D = 1.45$ μs. Upper left panel: excitation (blue line) and PL (red line) signal in time; upper right panel: amplitudes of Fourier components in excitation (blue line) and PL (red line); lower left panel: retrieved PL decay curve in a linear scale; lower right panel: retrieved PL decay (red line) compared to the original decay (blue line) in a semilogarithmic scale.

Fig. 2 also shows a simulated measurement of bi-exponential PL decay $I_D$ (see Eq. (6)) using the aforementioned modulated $I_{EXC}$ with a random distribution of duty cycle (random digital signal).

$$I_D = A_1 e^{-t/\tau_1} + A_2 e^{-t/\tau_2}. \tag{6}$$

Fig. 2, which shows the digital modulation, can be directly compared with Fig. 3, which presents the simulated measurement of the same bi-exponential PL decay $I_D$ using a random analog $I_{EXC}$ signal. In both figures, the parameters of the used bi-exponential $I_D$ were $A_1 = 1$, $A_2 = 15$, $\tau_1 = 40$ μs and $\tau_2 = 5$ μs. Impulse response function (IRF) for random digital signal was $IRF_D = 1.45$ μs and for random analog signal it was $IRF_A = 2.07$ μs.

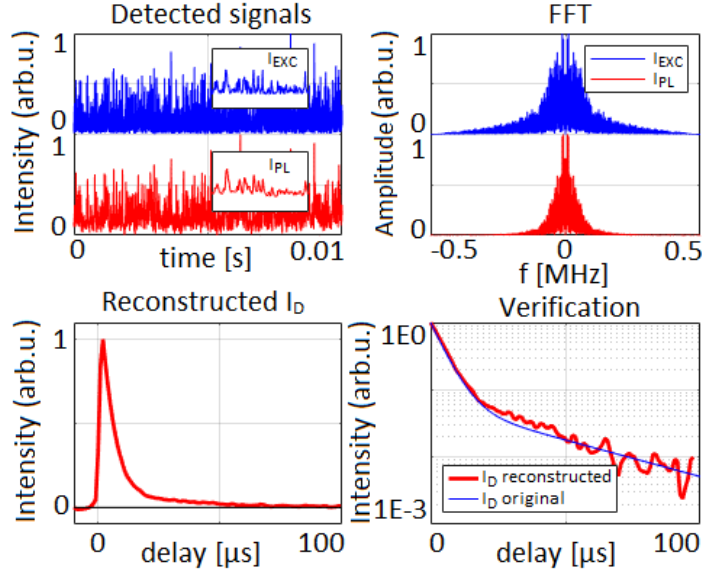

Fig. 3. Simulated measurement of bi-exponential $I_D$ with parameters $A_1 = 1$, $A_2 = 15$, $\tau_1 = 40$ μs and $\tau_2 = 5$ μs using analog random signal with $IRF_A = 2.07$ μs. Upper left panel: excitation (blue line) and PL (red line) signal in time; upper right panel: amplitudes of Fourier components in excitation (blue line) and PL (red line); lower left panel: retrieved PL decay curve in a linear scale; lower right panel: retrieved PL decay (red line) compared to the original decay (blue line) in a semilogarithmic scale.

By comparing the simulated PL decay retrieval for the digital and analog modulation, we can conclude, in agreement with our expectations, that both the approaches lead to identical PL decay. Therefore, digital modulation of excitation intensity can be used without losing the credibility of the retrieved PL decay. Although the spectrum of a digital signal features a narrower central peak in the Fourier space, it has higher amplitudes for higher frequencies. Therefore, the digital signal reaches a lower IRF width compared to the analog one.

We present the experimental confirmation of the agreement in Section 4. It is worth noting that the applicability of the analog modulation approach had been previously verified with a streak camera or TCSPC [21,23].

### 3.2 Direct PL lifetime map reconstruction

The original approach to PL lifetime imaging (FLIM) was based on the retrieval of a PL map for each delay after excitation, as it is introduced in Section 2.2. The real-life measurements showed that this procedure makes the post-processing routine highly time-consuming. Therefore, we propose an alternative approach that allows determining the FLIM spectrogram solely from the number of reconstructions, which equals the expected number of lifetimes in the measured sample. Namely, by determining a set of viable PL lifetimes in the sample, we can mathematically reconstruct the amplitude maps directly for the given lifetime $\tau$. This approach is highly beneficial for samples with a set of PL markers, mapping samples with distinct defects, or color centers emitting with well-defined lifetime.

In the cases where we know a priori the present lifetimes in a sample—for instance, a sample with a set of PL markers [28]—we can directly use this knowledge. However, in the opposite case, it is necessary to take the so-called zeroth step, i.e., illuminate the entire measured area of a sample with a homogeneous excitation spot and determine the PL decay curve $I_{DA0}$ representing the whole sample, using Eq. (4). The $I_{DA0}$ curve can be used to extract all present lifetimes via fitting. Since the entire spectrum of lifetimes is obtained from the zeroth step, it is then sufficient to fit only amplitudes of each lifetime during the $I_{DA}$ investigation (see text below). The literature states that it is appropriate to assume the fitting with a bi-exponential or tri-exponential decay [14]. Thus, we tested the novel principle of post-processing for bi-exponential decay (see Eq. (6)).

The direct PL lifetime map reconstruction is based on the single-pixel camera concept. Therefore, it can be used the same measurement routine. The measured sample was illuminated with random patterns (masks), which follow the temporal fluctuation of intensity according to $I_{EXC}$. The $I_{DA}$ corresponding to each mask was reconstructed according to Eq. (4). The obtained $I_{DA}$ curves were then fitted with a multi-exponential function with a fixed set of lifetimes. The fitting provided us with the amplitudes corresponding to the individual lifetimes.

If we consider the bi-exponential decay curve of the sample (see Eq. (6)), we obtain the parameters $\tau_1$ and $\tau_2$ from the fitted $I_{DA0}$. We assume the presence of $\tau_1$, $\tau_2$ in all measured $I_{DA}$ curves and, therefore, we only fit parameters $A_1$ and $A_2$ of all $I_{DA}$ curves. Because the sample is illuminated by the number of masks $M$ (given by the compression ratio $k$), we also get $M$ different values of $A_1$ and $A_2$, which creates vectors $A_1 SPC$, $A_2 SPC$ of size $M$. Since the amplitudes $A_1$ and $A_2$ are a linear superposition of all PL decays within the illuminated area of the sample, we can employ the same retrieval algorithms as we use for the PL intensity map.

The knowledge of the used masks and $A_1 SPC$, $A_2 SPC$ vectors can be used to reconstruct amplitude maps using Eq. (7). In this case of bi-exponential decay, we obtain two amplitude maps of lifetime $H\tau_1$ and $H\tau_2$.

$$min\left\{\left\|BH_{\tau n} - A_n SPC\right\|_2^2 + TV(H_{\tau n})\right\}. \tag{7}$$

In the case of monoexponential decay, the reconstructed map $H_\tau$ represents the amplitude distribution for a given $\tau$. Nevertheless, if it is an $n$-exponential decay, the lifetime $\tau$ in each pixel (FLIM spectrogram) must be determined. There are multiple approaches to calculating the overall decay lifetime of a complex PL decay curve. Here we consistently used the weighted average:

$$\tau(x, y) = \frac{\sum_{i=1}^{n} H_{\tau i}\tau_i}{\sum_{i=1}^{n} H_{\tau i}}. \tag{8}$$

The process of amplitude map retrieval is described in more detail in a scheme assuming the bi-exponential decay ($\tau_1 = 20$ ns, $\tau_2 = 70$ ns), as shown in Fig. 4. Amplitude distribution of $\tau_1$ can be observed in $H\tau_1$ map, and amplitude distribution of $\tau_2$ is shown in $H\tau_2$ map. Using Eq. (8) we calculated the total FLIM spectrogram with the mean lifetime.

The proposed approach significantly accelerated the post-processing procedure, and the whole FLIM spectrogram can be determined using $n$ reconstructions for $n$-exponential decay. Thus, in the case of bi-exponential decay, there are only two reconstructions of the undetermined system. Moreover, this approach offers excellent resistance to noise in the measured data, as shown in the next subsection.

Due to its principles, the presented direct lifetime retrieval is more suited for materials with several emitting species with sharp PL lifetime values, where the contribution of each species varies over the scene. In such a case, the method outperforms the standard approach, as we will show later. In the case where each pixel of the scene features a different lifetime value from a broad distribution, the method can still attain reasonable results by including more lifetimes and calculating the mean PL lifetime according to Eq. (8). Nevertheless, for a very broad lifetime distribution, using the standard frame-by-frame retrieval can be considered as a better option.

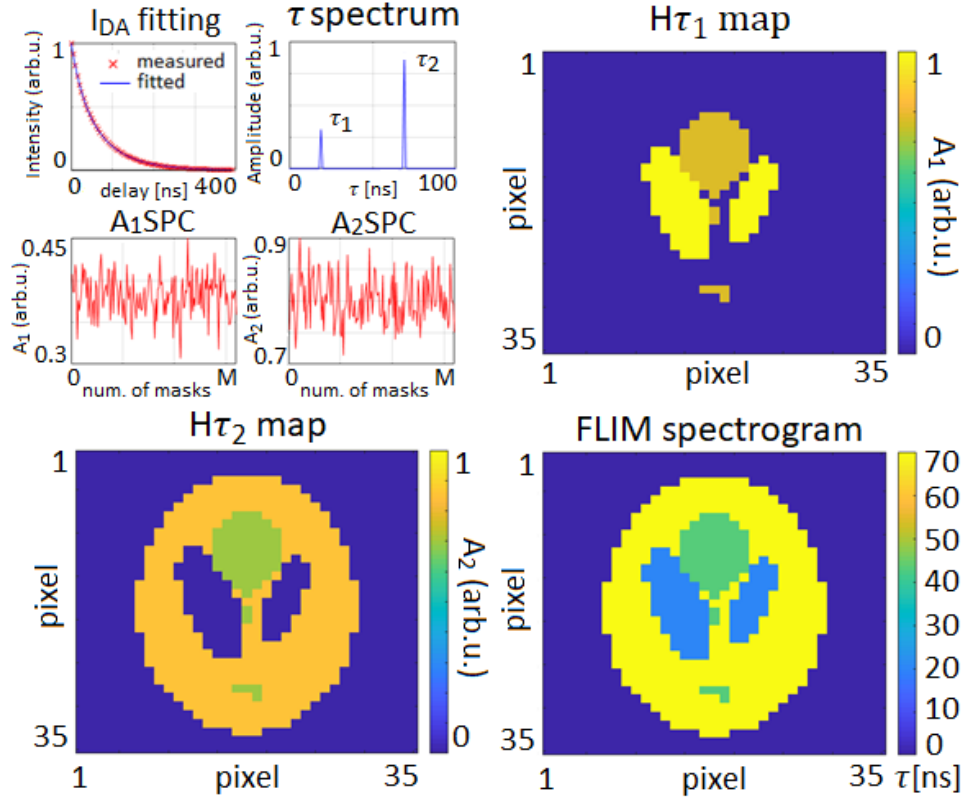

Fig. 4. (A) Example of fitting $I_{DA}$ curve amplitudes $A_1$ and $A_2$, where the distribution of $\tau_1$ and $\tau_2$ is already known from $I_{DA0}$ fitting. The amplitude fitting provides vectors $A_1SPC$ and $A_2SPC$. (B) Reconstructed amplitude map $H\tau_1$. (C) Reconstructed amplitude map $H\tau_2$. (D) Calculated FLIM spectrogram based of knowledge of $H\tau_1$ and $H\tau_2$.

*3.3 The comparison of noise effect on both post-processing approaches*

The new approach to the FLIM spectrogram evaluation introduced in Section 3.2 significantly reduces the time required for post-processing compared to the original approach described in Section 2.2. In this section, we compare both approaches with respect to the quality of the obtained FLIM spectrogram with different levels of noise present in the measured data.

In our previous work, we showed that the quality of the retrieved $I_{DA}$ and the 2D scene reconstruction is significantly more affected by the noise level present in $I_{PL}$ compared to the $I_{EXC}$. Furthermore, it was shown that the quality of 2D scene reconstruction is negligibly affected by the change in compression ratio compared to the amount of noise in the system [25]. Therefore, for the following simulations, the compression ratio was always kept at $k = 0.4$, and we tested only the noise present in the PL signal $I_{PL}$, which we set to 0%, 0.5%, 1%, and 1.5%, respectively. The $I_{EXC}$ signal was kept noiseless. The resulting FLIM spectrogram of lifetimes $F$ was always compared with the ground truth $U$ to extract the percentage error of the reconstructed image $R$:

$$R = \frac{\sum \sqrt{(F-U)^2}}{\sum \sqrt{U^2}} \cdot 100. \tag{9}$$

The spectrograms obtained by the original approach described in Section 2.2 are labeled as FLIM$_B$, while the spectrograms obtained by the new, proposed approach are labeled as FLIM$_A$. The simulation results are summarized in Fig. 5, where the individual row represents the different noise levels in the $I_{PL}$ signal, while the columns indicate gradually the $H\tau_1$, $H\tau_2$, FLIM$_A$ and compare them to the FLIM$_B$. The dependence of the error $R$ on the noise level is then provided in Fig. 6.

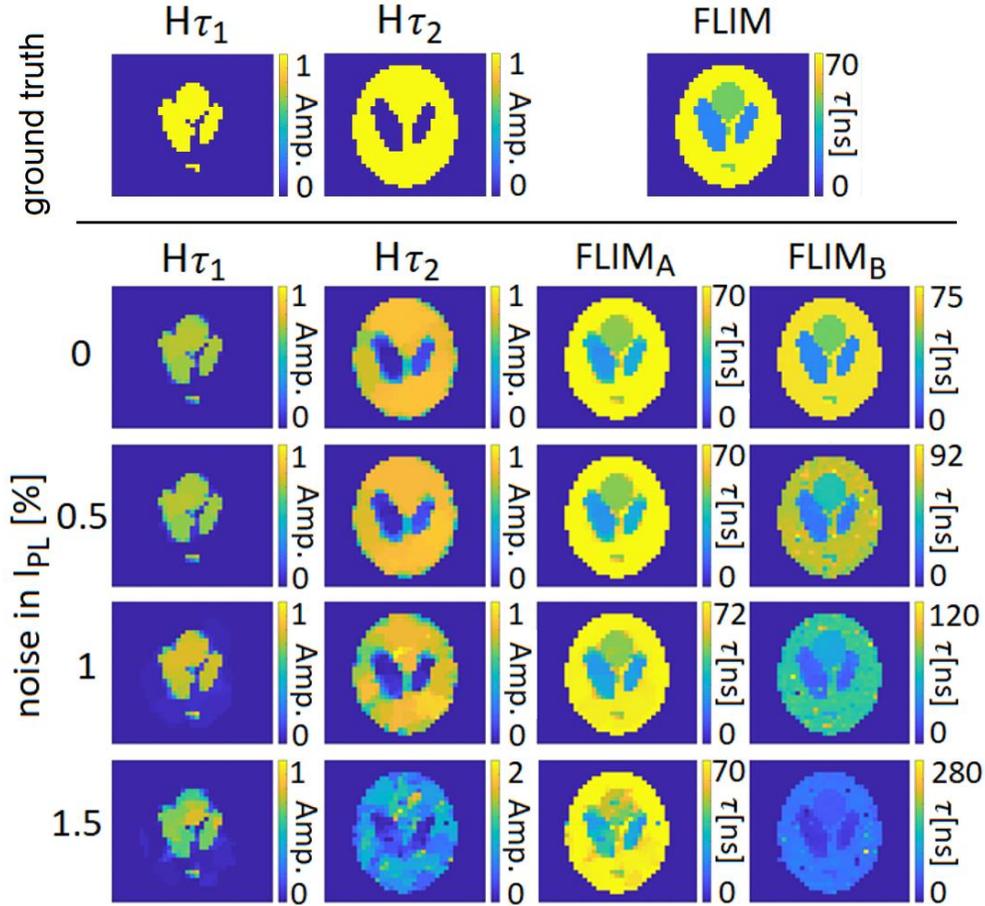

Fig. 5. Above the line is the ground truth, which can be compared to the simulated results below the line. The results of simulations of noise effect on both post-processing approaches. The new approach, represented through FLIM$_A$ (the third column), is supplemented by reconstructions of partial amplitude maps $H\tau_1$ (the first column) and $H\tau_2$ (the second column). The results of the original approach are represented through FLIM$_B$ (the fourth column). Each row corresponds to chosen noise level in the system (0-1.5%).

For this simulation, a scene with two areas $\tau_1 = 20$ ns and $\tau_2 = 70$ ns and corresponding amplitudes $A_1 = 1$ and $A_2 = 1$ was employed. The "sample" combines parts with a mono-exponential decay with a region featuring a bi-exponential decay.

The TVAL3 algorithm was used to reconstruct the undetermined systems in Eq. (7) [29,30]. In the case of FLIM$_B$, it was the intensity PL maps reconstruction, where the main parameters of the algorithm were *mu* ($2^{11}$) and *beta* ($2^7$). In the case of FLIM$_A$, amplitude maps reconstruction was used, where the main parameters were set to *mu* ($2^9$), and *beta* ($2^6$). The mentioned parameters *mu* and *beta* were optimized for both approaches independently to attain the best reconstruction results with the lowest residues.

Reconstructed pixels that were below 10% of the PL amplitude were removed from the statistics for both FLIM$_A$ and FLIM$_B$ in Fig. 5 and also in Fig. 6. The fitting curves for both FLIM$_A$ ($I_{DA}$'s) and FLIM$_B$ (3D-datacube) cases were always in the range of 0 ns to 400 ns, which is sufficient for the entered parameters $\tau_1$ and $\tau_2$.

For a rigorous discussion of results, it is important to realize that the FLIM$_B$ results (original approach) are dominantly affected by noise in $I_{SPC}$ (see Fig. 1(B)). This noise arises from inaccuracies in the reconstruction of individual PL maps $m(x,y,t)$ when creating a 3D datacube (see Fig.1(C)). Due to the character of the image retrieval, the noisy data entering the fit can cause a vast local error of the PL lifetime value in a few pixels. Therefore, FLIM$_B$ spectrograms could deviate from the ground truth values locally, while the overall agreement might be kept.

On the contrary, the FLIM$_A$ approach could introduce an error already in the zeroth-step of the process, when the whole sample is homogeneously illuminated, and the set of lifetimes contained in the sample is analyzed via $I_{DA0}$ fitting. The obtained lifetimes from the zeroth step are further considered during the remaining $I_{DA}$ evaluation, where they are used to calculate the mean PL lifetime via amplitudes $A_1$ and $A_2$. Since the amplitude maps are calculated from an undetermined system with the condition of a low total image variation, we attained smooth images without local errors despite the present noise. On the other hand, the error of the initial lifetime fit propagates to the final FLIM spectrogram. Therefore, in our simulations, the reconstruction of 0% noise in the system, FLIM$_B$ (original FLIM analysis) has higher accuracy than FLIM$_A$. However, in a situation with a PL noise of 0.5% and higher, FLIM$_A$ is more accurate than FLIM$_B$ (see Fig.6).

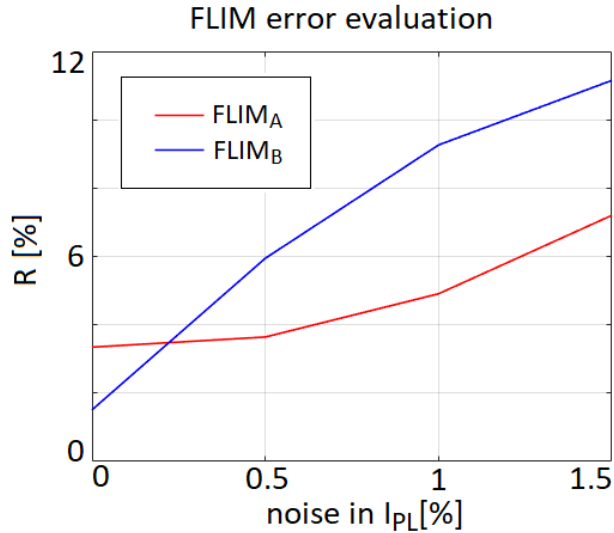

Fig. 6. Evaluation of FLIM spectrogram reconstruction error – see Eq. (9) - via FLIM$_A$ (direct PL lifetime map) and FLIM$_B$ (frame-by-frame standard analysis) approach for a different amount of noise in a system ($I_{PL}$ signal).

Although it may seem that an imprecise fit of PL lifetimes in the zeroth step will highly affect the FLIM$_A$ results, our tests proved that this effect is not pronounced. We summarized in Table 1 all the obtained lifetimes after the zeroth step ($I_{DA0}$ fitting), which all suffer from the inevitable fitting error. Nevertheless, the total error of the FLIM$_A$ spectrograms—see Fig. 6—increases with the noise in the fitted data and does not follow the lifetime errors. As an example, the worst zeroth-step lifetime fit is associated with the noise of 1% in the system (see Table 1), while the highest total reconstruction error is linked to the data with the noise of 1.5% (see Fig.6). To outline the fitting issue, we present results obtained after a single $I_{DA0}$ trace fitting. However, the imprecise determination of lifetimes can be improved by statistical averaging over several measurements in the zeroth step.

Table 1. Summary of results of $I_{DA0}$ fitting in zeroth step of FLIM$_A$ determination process.

| Noise in I$_{PL}$ signal [%] | $\tau_1$ [ns] | $\tau_2$ [ns] |
|---|---|---|
| 0 | 21 | 70 |
| 0.5 | 21 | 70 |
| 1 | 24 | 72 |
| 1.5 | 22 | 70 |

Overall, it can be stated that the direct PL lifetime map reconstruction (FLIM$_A$) is more stable in terms of noise. The standard frame-by-frame retrieval (FLIM$_B$) may be locally accurate, while some pixels may be entirely out of scale, and it is not easy to determine whether this is an effect of the noise of an actual sample property.

It is beneficial to address new methods for accurately determining the parameters of exponential curves [31], which would further improve the quality of both considered approaches. Nevertheless, the novel presented approach is less time-consuming in post-processing than the standard one.

## 4. Experimental implementation

### 4.1 Optical setup

The advanced optical setup stemmed from the original proof-of-principle experiments [23]. Its scheme is shown in Fig. 7.

Temporal modulation was ensured via modulated Cobolt laser S06-01 MLD (405nm). The modulation signal was produced by the Digilent Cmod A7 development kit and was generated in the FPGA Xilinx Artix-7 (VIVADO software package). The bitstream is generated via Linear Feedback Shift Register (LFSR) from flip-flops and XNOR gate feedback, configured in FPGA. The output of the LFSR meets many randomness tests [32]. Using the above-mentioned configuration, one can generate any bitstream with the desired repetition period of the random signal. Thus, it is possible to choose as long a period as necessary for one measurement and thus avoid the effect of $I_{EXC}$ periodicity described in our previous work [25].

A temporally modulated laser beam is expanded using a GBE20-A Thorlabs beam expander and guided to the digital micromirror device (DMD) through mirrors M$_1$ and M$_2$ so that one of the reflected beams is reflected in the normal direction of the DMD chip surface. In order to achieve an even distribution of the DMD-generated illumination pattern, the laser beam diameter is magnified 20 times in front of the DMD, so it significantly exceeds the dimensions of the DMD chip.

The DLPLCR65EVM DMD from Texas Instruments was controlled via a DLPLCRC900EVM module. We used 35×35 pixel patterns both in the simulations (Section 3) and in the demonstration measurements (Section 4). However, rather than the entire DMD chip size, only a limited part of the chip in its center was used to generate patterns. This enabled us to use a low numerical aperture within the setup.

Because the DMD pixel size is in the micrometer scale (7.56 μm), diffraction on the pixels should be considered [33,34]. Therefore, the diffracted beam from the DMD is collected with an AR-coated fused silica lens $L_1$ (Ø50mm, f = 75 mm) and then filtered using a low pass filter in the configuration of $L_1$ and the second AR-coated fused silica lens $L_2$ (Ø25mm, f = 50 mm) and iris aperture. Hence, only one diffraction order passes through the low pass filter and hits the Thorlabs N-BK7 beam splitter BS (CM1-BS013 (50:50)). The reflected part of the beam is targeted on a PDA10A2 photodiode for $I_{EXC}$ detection, while the transmitted part of the beam goes through a 405 nm bandpass filter $F_1$ (Thorlabs FBH405-10) to avoid possible parasite $I_{PL}$ generation in N-BK7 glass (BS).

Subsequently, the beam is transmitted through a dichroic mirror DM (Thorlabs DMSP425, 425 nm cut-off) and imaged by a microscope objective—4X Olympus plan achromat objective (0.10 NA, 18.5 mm WD). Using the 4X Olympus plan achromat lens, we ideally get a single pixel size of 12.6 μm and a field of view of about 450x450 μm. The stated values were confirmed with calibration grids with a defined line spacing of 50 μm.

The illumination pattern imaged by a microscopic lens on a sample can be captured using a CMOS camera (IDS UI-3240ML-M-GL) to find the optimal focus. The microscopic lens also collects the generated $I_{PL}$ signal, which is reflected and focused on a photomultiplier (Hamamatsu H10721-20) through a dichroic mirror and lens $L_3$ (Ø25mm, f = 12 mm). Prior to PL detection, the collected light is filtered by the Thorlabs 500 nm cut-off filter $F_2$ (FELH0500) to avoid a possible back reflection of $I_{EXC}$. Thus, all decay curves correspond to PL in a range above 500 nm of wavelength.

The temporal waveforms of both the $I_{EXC}$ and the $I_{PL}$ signals are collected by using the TiePie Handyscope HS6-1000XM. The impulse response function (IRF) depends on the modulation frequency and many additional parameters and we provide IRF width for each experiment.

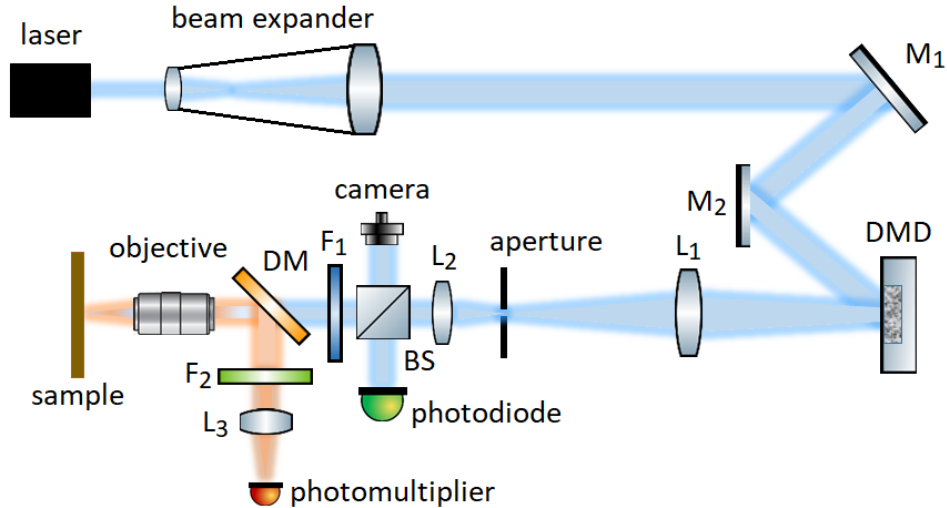

Fig. 7. The 2D-RATS method in a single-pixel camera configuration implemented into a microscope optical setup.

In the current configuration, where we used a cw-operated laser at 125 mW, we attained an intensity of the excitation light of 0.35 mW at the sample plane. This means that the setup efficiency was about 0.3%. Note that the use of excitation laser efficiency was not optimized, and it was highly affected by using an excessive beam diameter at the DMD, the use of a part of the DMD chip, selection of a single diffraction order, or beam splitter ratio—these parameters can be tuned to improve efficiency. Yet, we reached efficiency a hundred times higher compared to the proof-of-principle diffusor-based 2D RATS setup (efficiency of about 0.003%) [23].

*4.2 Digital and analog modulation of the excitation signal*

By comparing the simulated PL decay retrieval for the digital and analog modulation in Section 3.1, we concluded, in agreement with our expectations, that both the approaches lead to identical PL decay. We used the new optical setup to carry out the same comparison experimentally.

Both approaches (analog vs. digital random $I_{EXC}$) were compared in a single-point measurement (0D-RATS) configuration, where the SCHOTT OG565 absorption filter was used as a test sample. The setup described in this article was used for the digital excitation modulation, while the proof-of-principle setup was employed for the analog excitation modulation [21]. For the analog modulation, we reached an improved analog modulation speed by tightly focusing an expanded excitation beam (f = 25 mm, spot size 2.5 µm) on a fine diffusor produced with SiC1200 abrasives. Both the beam distance from the diffuser center of 115 mm and the rotation speed of 100 Hz contributed to the fast modulation. Therefore, the IRF width for the analog case was $w_A = 71$ ns. The digital modulation was adjusted to reach a similar IRF width of $w_D = 47$ ns.

To gain a rigorous comparison, the $I^0_D$ curve obtained using the random digital signal was convolved with the Gaussian function $G(w)$ with a FWHM $w$ equal to the root mean square difference of $w_A$ and $w_D$ values:

$$I_D = I_D^0 * G\left(\sqrt{w_A^2 - w_D^2}\right). \quad (10)$$

The experimental results in a semilogarithmic representation are presented in Fig. 8. They confirmed that the digital and analog modulation of the signal can be equally employed in the PL decay analysis and lead to the same decay functions.

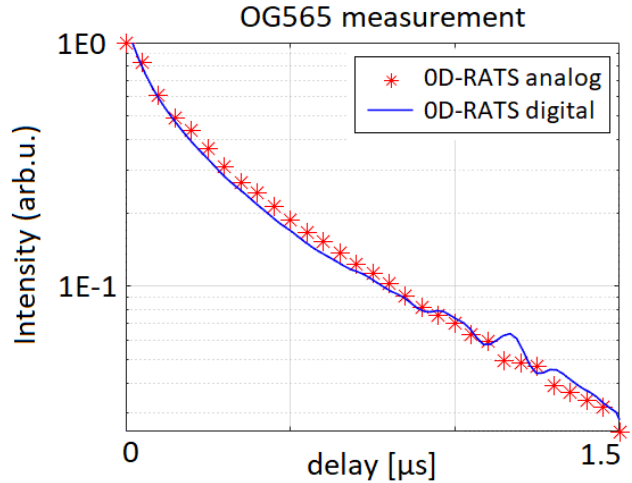

Fig. 8. Comparation measurement of OG565 filter using analog and digital random $I_{EXC}$ with $w_A = 71$ ns and $w_D = 47$ ns.

*4.3 Temporal resolution of digital modulation of the excitation signal*

By using the thorough optimization of the diffuser-based intensity modulation, we reached the IRF width of 71 ns, as stated before. Now, we will turn to show the potential of digital modulation using the Cobolt S06-01 laser. Here, we tested the variation of both the laser modulation frequency and the sampling rate—see Fig. 9. We can see that the fastest combination leads to the IRF width of 6 ns, which is an order of magnitude below the optimized diffuser-based modulation.

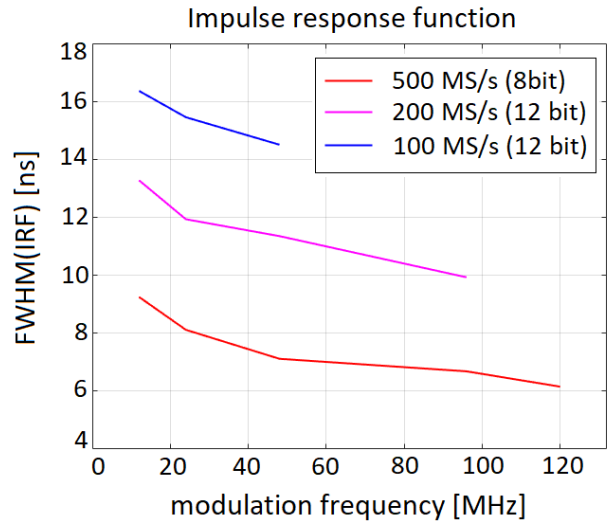

Fig. 9. IRF measurement for different modulation frequencies and different sampling frequencies.

That is primarily due to the laser parameters, especially modulation. Hence, it is easy to generate a much faster random signal than using our previously described rotary diffuser-based generator [21]. Apart from this, a prominent feature in Fig. 9 is that the IRF width does not strongly depend on the modulation frequency. This behaviour is the consequence of the rectangular character of the digital signal. From the general principles of rectangular signal generation, the signal contains rapidly rising and falling edges carrying significantly higher frequencies than the set modulation frequency of the digital signal. These frequencies would be revealed with sufficient sampling and bandwidth. Therefore, ideally, a digitizer with significantly faster sampling and a broader bandwidth should be used for such a (i.e., modulated) signal.

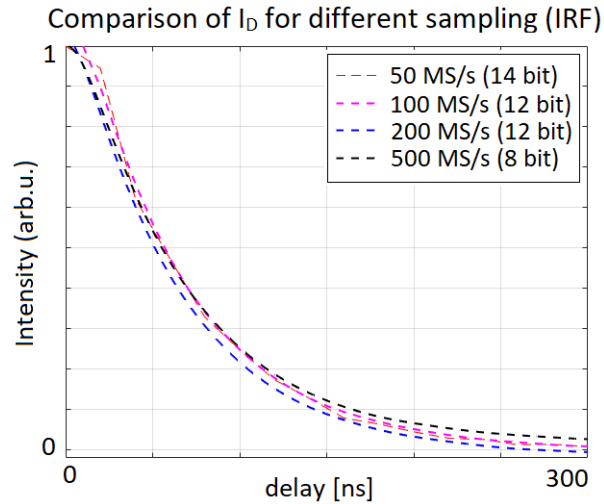

Fig. 10. Comparison of $I_D$ measured on LuAG:Ce for different sampling frequencies (different IRF width). Measurements with lower IRF width were convolved with the root square deviation of IRF width related to measurements with the highest IRF width.

The employed HS6-1000XM TiePie USB-handyscope allows sampling of up to 500 MSa/s with a bandwidth of 250 MHz when scanning on two channels simultaneously. Hence the

Nyquist condition for the fastest bit carrier frequency was met, but the frequencies of leading and trailing edges were truncated. As a result, the IRF width strongly depends on the sampling rate, as shown in Fig. 9.

However, since the carrier frequencies are not truncated, no aliasing occurs, and thus the resulting $I_D$ reconstruction is not distorted. To confirm this statement, we plot in Fig. 10 PL decay of the LuAG:Ce luminophore for various sampling options. As in Fig. 9, each measurement was convolved with a Gaussian function to match the largest IRF width within the measurements. This allows for a direct comparison of the curves. We did not observe any notable variation in the decay curves and corresponding PL lifetime for the sampling rates well above the PL decay rates.

In total, a RATS setup with digital signal modulation, which is not limited by the bandwidth of the detector and the corresponding electronics, has the ultimate resolution due to the steepness of the rectangular laser signal edge and the sampling rate.

### 4.4 Demonstration measurement

The 2D-RATS experiment based on the advancements presented in Section 3.2 was experimentally implemented into a microscope setup based on a single-pixel camera configuration. The technical details of the setup are described in Section 4.1. Here we provide an example of data acquired by the advanced setup and data processing.

We selected a LuAG:Ce scintillation crystal as a test sample [35], which features strong PL emission. The test sample was a thin monocrystal plate polished on both sides with visible, microscopic scratches and cracks. One of the cracks was selected as the testing spot since we expected a significant increase in PL intensity at the defect. The PL intensity is higher due to high scattering, which improves PL signal outcoupling from the sample.

A 4X Olympus plan achromat objective (0.10 NA, 18.5 mm WD) was used to map the testing area of 450×450 μm with a pixel size of 12.6 μm (35×35 pixels). The compression ratio was set to k = 0.4 (490 measurements) with a single PL decay dataset measured over 10 ms. However, the current total acquisition time in the order of tens of minutes was caused dominantly by the data transfer and data handling, which can be significantly optimized. The sampling frequency was 200 MHz and the modulation frequency of the fastest bit of the random signal reached 98 MHz, which resulted in IRF width of 9.9 ns.

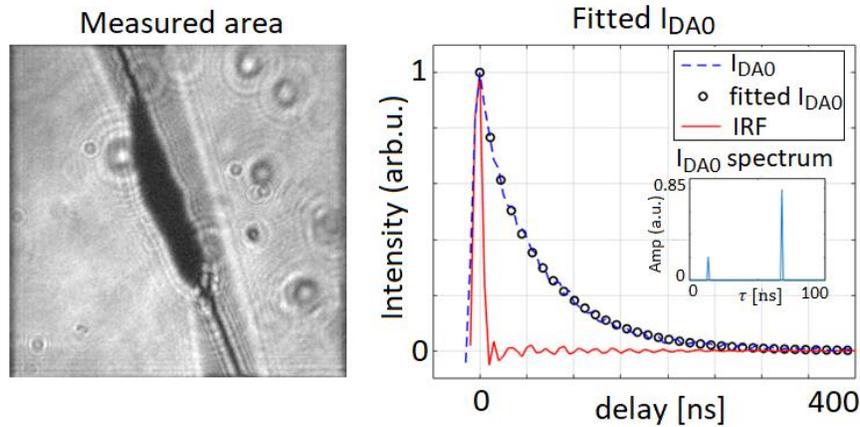

Fig. 11. Left panel: Measured area of the LuAG:Ce crystal. Right panel: Corresponding $I_{DA0}$ with fitted bi-exponential curve and revealed spectrum of the $\tau_1$ and $\tau_2$.

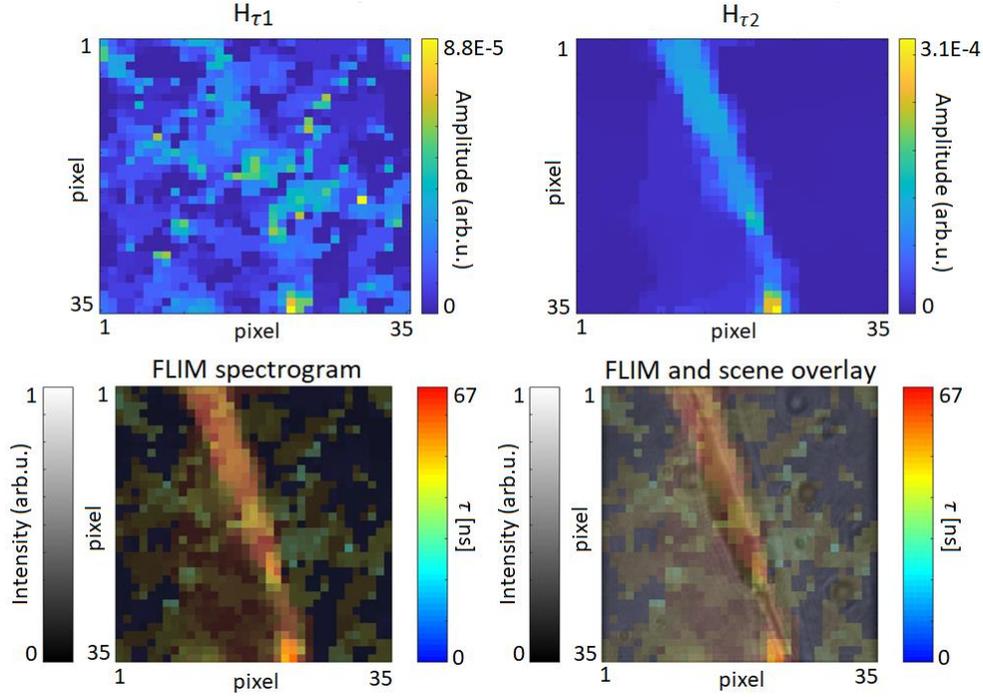

Fig. 12. Upper panels: Reconstruction of $H\tau_1$ and $H\tau_2$ areas corresponding to expected lifetimes.
Lower panels: FLIM spectrogram determination and its overlay with the measured scene.

In the initial analysis, we tested the measurement for a more general bi-exponential fit, even though we expected the LuAG:Ce to show a monoexponential decay. The mapped region and the corresponding $I_{DA0}$, together with the IRF, are shown in Fig. 11. Fitting the $I_{DA0}$ revealed the PL lifetimes of $\tau_1 = 11$ ns and $\tau_2 = 67$ ns.

The results of mapping $H\tau_1$, $H\tau_2$, and the FLIM spectrogram are presented in Fig. 12. The map of the first component can be assigned to random scattering points and scattered light along the crack. The map of the second component closely follows the shape of the crack, where the PL from LuAG:Ce is efficiently coupled out from the crystal. The FLIM spectrogram combines both amplitudes, where the second component dominates, owing to its high amplitude. The brightness of the individual pixels in the FLIM spectrograms was scaled according to the PL intensity, i.e., the sum of the amplitudes $A_1$ and $A_2$ in a given pixel. In Fig. 12, it is also possible to observe the overlay of the measured area and the obtained FLIM spectrogram. The overlay documents the agreement between the PL mapping and the sample properties. It is worth noting that only pixels with PL intensities greater than and including 10% of the maximum are shown in the FLIM spectrogram.

As we stated before, the assumption of the bi-exponential decay of LuAG:Ce was not physically correct. Moreover, $\tau_1$ is close to the IRF width, which distorts $A_1SPC$. Therefore, we also provide an analysis of the PL, where we assume a monoexponential decay. Such a situation would, for instance, correspond to the mapping of a biological sample with a single PL marker. For the monoexponential decay, we obtained only one $H\tau$ region for the lifetime of 59 ns, which corresponds well with the PL lifetime of LuAG:Ce [36]. The $H\tau$ is directly represented in the FLIM spectrogram, as shown in Fig. 13. We can see that in this case we attained a clear image of the crack with high PL intensity together with low-intensity PL regions surrounding the cracks where the PL is outcoupled from the monocrystal inefficiently.

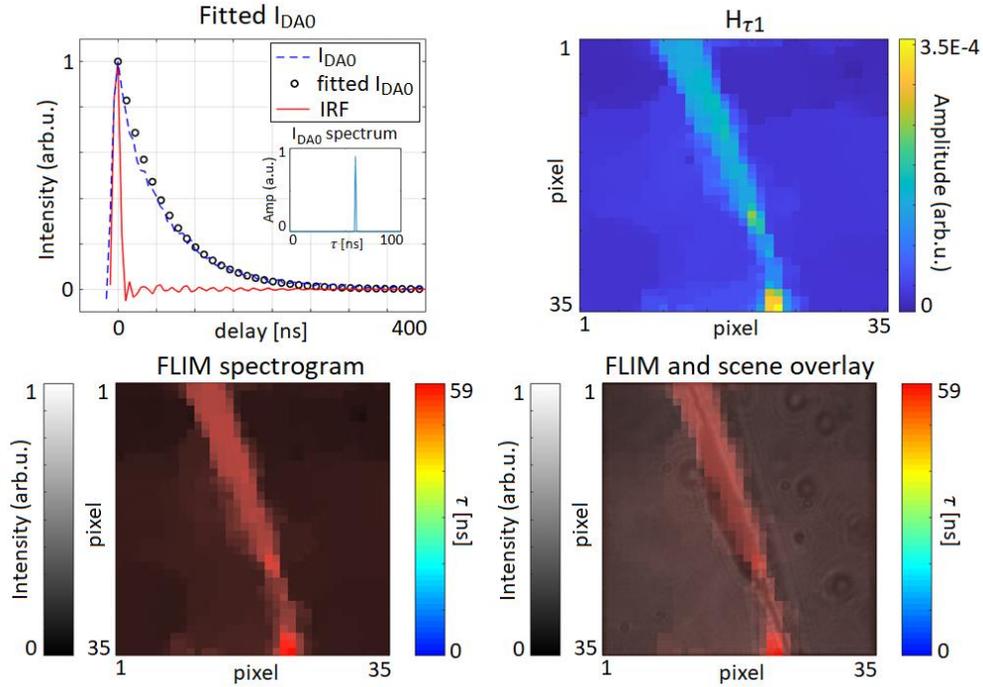

Figure 13: Upper panels: Fitted $I_{DA0}$ with monoexponential curve and reconstructed area $H\tau_1$ for corresponding $\tau_1$. Lower panels: FLIM spectrogram determination and its overlay with the measured scene.

## 5. Conclusion

We outline two significant advancements for the RATS method for measuring PL dynamics and FLIM. The first is the possibility of random excitation via a digitally modulated signal. Using this approach, it is possible to achieve a time resolution down to units of nanoseconds and to significantly simplify the optical setup. Secondly, the article also presents a new approach to the evaluation of the FLIM spectrogram, which significantly reduces the number of necessary reconstructions of the undetermined system and reduces the post-processing time accordingly. Moreover, the novel approach to data processing reduces the required number of fitted curves in proportion to the chosen compression ratio $k$.

The original and the newly proposed approach to FLIM retrieval were compared in simulations regarding noise analysis. The new approach was demonstrated to reduce the possibility of locally incorrect lifetime determination. Therefore, the proposed amplitude map retrieval is more robust compared to the standard analysis against increasing noise level.

The advancements were implemented in a microscope setup based on the single-pixel camera technique. A digital micromirror device (DMD) was used to generate random patterns (masks), which also enables full illumination of the scene to determine lifetimes within the sample. In addition, the combination of DMD and random digital modulation of the laser increases the efficiency of the optical system a hundred times compared to the original arrangement using diffusers [23].

We analyzed the advancements on synthetic data, as well as on testing measurements. Namely, we carried out the imaging of a LuAG:Ce crystal, where a crack in the crystal was mapped. The resulting FLIM spectrograms from the PL analysis were in perfect agreement with the camera images.

**Funding**

This work was supported by the Ministry of Education, Youth and Sports ("Partnership for Excellence in Superprecise Optics," Reg. No. CZ.02.1.01/0.0/0.0/16_026/0008390), the Czech Science Foundation (GACR) (Grant number: 22-09296S), and the Student Grant Competition at the Technical University of Liberec (under project no. SGS-2021-3003).

**Acknowledgment**

We gratefully acknowledge CRYTUR company for providing us with LuAG:Ce crystals, and Jakub Nečásek for his help with a digital modulation device.

**Disclosure**

The authors declare no conflicts of interest.

**Data availability**

Data underlying the results presented in this paper are not publicly available at this time but may be obtained from the authors upon reasonable request.